\documentclass[letterpaper, 10 pt, conference]{ieeeconf}  
\IEEEoverridecommandlockouts                             
\usepackage[english]{babel}
\usepackage{amsmath, amssymb, amsfonts, bm}
\usepackage{mathrsfs}
\usepackage{textcomp}
\usepackage[mathcal]{euscript}
\usepackage{enumerate}
\usepackage{mathbbol}
\usepackage{float}
\usepackage{color,graphicx,epstopdf}
\usepackage{dsfont}
\usepackage{hyperref}
\usepackage[noadjust]{cite}
\usepackage{caption}
\usepackage{subcaption}
\usepackage{mathtools}
\usepackage{tikz-cd}
\usepackage{mathtools}
\usepackage{algorithm}
\usepackage{algpseudocode}
\usepackage{url}
\usepackage{authblk}
\usepackage{verbatim}
\usepackage{stackengine}
\usepackage{subcaption}
\usepackage{xcolor}
\usepackage[thinlines]{easytable}

\newtheorem{theorem}{Theorem}
\newtheorem{definition}{Definition}

\newtheorem{lemma}{Lemma}
\newtheorem{remark}{Remark}

\newtheorem{assumption}{Assumption}

\usepackage[left=1.91cm,right=1.91cm,top=1.91cm,bottom=1.91cm]{geometry}

\renewcommand{\color}[1]{}

\newcommand{\sat}{{\text{sat}}}

\title{\LARGE \bf
	A Nonlinear Negative Imaginary Systems Framework with \\Actuator Saturation for Control of Electrical Power Systems
}

\author{Yijun Chen$^{1}$, Kanghong Shi$^{1}$, Ian R. Petersen$^{1}$, and Elizabeth L. Ratnam$^{1}$
	
	\thanks{$^{1}$The  School of Engineering, The Australian National University, Canberra, Australia, emails: \{yijun.chen, kanghong.shi, ian.petersen, elizabeth.ratnam \}@anu.edu.au.}%
	
	\thanks{This work was supported by the Australian Research Council under grants DP230102443 and LP210200473.}
}

\begin{document}

	\maketitle
	\thispagestyle{empty}
	\pagestyle{empty}
	
	\begin{abstract}
	In the transition to net zero, it has been suggested that a massive expansion of the electric power grid will be required to support emerging renewable energy zones. In this paper, we propose the use of battery-based feedback control and nonlinear negative imaginary systems theory to reduce the need for such an expansion by enabling the more complete utilization of existing grid infrastructure. By constructing a novel Lur\'e-Postnikov-like Lyapunov function, a stability result is developed for the feedback interconnection of a nonlinear negative imaginary system and a nonlinear negative imaginary controller. Additionally, a new class of nonlinear negative imaginary controllers is proposed to deal with actuator saturation. We show that in this control framework, the controller eventually leaves the saturation boundary, and the feedback system is locally stable in the sense of Lyapunov. This provides theoretical support for the application of battery-based control in electrical power systems. Validation through simulation results for single-machine-infinite-bus power systems supports our results. Our approach has the potential to enable a transmission line to operate at its maximum power capacity, as stability robustness is ensured by the use of a feedback controller.
	\end{abstract}
	
	\section{Introduction}	
	
	To meet the global challenge of climate change and the rapid transition to net-zero electricity, it has been suggested that a massive expansion of the electric power grid will be required to support emerging renewable energy zones  and the electrification of
	other sectors including transportation \cite{AEMO_TE}. In this paper, we propose the use of battery-based feedback control and nonlinear negative imaginary systems (NNI) theory \cite{lanzon2008stability,ghallab2022negative,shi2023output} to reduce the need for such an expansion by enabling the more complete utilization of existing grid infrastructure.
	
	Frequency deviations in electric power systems pose risks such as equipment damage, transmission line overloads, and compromised system reliability \cite{bevrani2014robust}. Therefore, the primary aim of frequency control is to maintain synchronization with the nominal frequency. Regulating frequency involves managing both the generation and load sides. Automatic generation control, a conventional frequency regulation scheme on the generation side, adjusts generator setpoints in response to frequency variations and unexpected power flows between different areas \cite{bevrani2014robust,glover2012power}. However, with the increasing integration of renewable energy sources, the heightened volatility in non-dispatchable renewable generation poses challenges for generator-side control, shifting attention towards load control strategies.
	
	In recent years, advancements in battery technologies have led to the widespread adoption of rechargeable batteries in electric vehicles, the use of large grid storage batteries, and the proliferation of domestic solar-powered batteries \cite{tran2019efficient,borenstein2022s}. In addition to energy storage supporting local power management, this transformative development may also open the possibility for energy storage systems to actively participate in regulating system frequency and ensuring transient stability in power systems. Nevertheless, integrating distributed energy storage systems into power systems necessitates a profound understanding of the interaction between rapidly controllable batteries and the dynamic behavior of power systems. Considering the possibility of real-time angle measurements for future power systems \cite{vivsic2020synchronous}, the work of \cite{chen2023design} proposes angle feedback linearization control for power transmission networks using linear negative imaginary (NI) systems theory \cite{petersen2010feedback,wang2015robust,xiong2010negative} to guarantee transient stability and frequency synchronization. However, it does not take into account actuator saturation, which is of a nonlinear nature. For this reason, it is useful to investigate controllers with actuator saturation in a nonlinear systems framework. In addition, NNI theory \cite{lanzon2008stability,ghallab2022negative,shi2023output} can be utilized to guarantee power system transient stability, ensure power system robustness, and address consensus problems relating to the convergence of bus angles.
	
	Therefore, this paper proposes a NNI systems framework with actuator saturation for control of electrical power systems. From the technical point of view, our contributions are two-fold: 1) we constructed a novel candidate Lyapunov function, fashioned in a manner reminiscent of the Lur\'e-Postnikov form \cite{haddad2008nonlinear} and intended for stability proofs for feedback interconnection of nonlinear negative imaginary systems; 2) we propose a new class of NNI controllers to deal with actuator saturation and ensure system stability. Moreover, our technical results provide theoretical support for applications in the field of electrical power systems. Our contributions to the area of electrical power systems are two-fold: 1) in the absence of a controller, we analyze the invariant set of the single-machine-infinite-bus (SMIB) power systems, which serves as a theoretical foundation for the stability condition arising from the well-known equal area criterion \cite{glover2012power}; 2) we present a control framework using real-time angle sensors and large-scale batteries as actuators with saturation, which exhibits the ability to enhance power system transient stability without the need for over-engineering.
	
	This paper is organized as follows: Section \ref{sec:preliminary} provides preliminary knowledge on Lyapunov's direct method for local stability and the definition of NNI systems. Section \ref{sec:NI_Feedback} provides the stability results for the feedback interconnection of NNI systems. Section \ref{sec:saturated_theory} proposes a NNI systems framework with actuator saturation for control. Section \ref{sec:SMIB} introduces SMIB systems. Section \ref{sec:local_storage} presents the application of the NNI systems framework with actuator saturation to the control of SMIB systems. In Section \ref{sec:simulation}, simulation results are given to illustrate our theory. Section \ref{sec:conclusion} concludes the paper.

	\section{Preliminaries}~\label{sec:preliminary}
	In this section, we recall Lyapunov's direct method for local stability and the definition of  {\color{red}NNI} systems.
	\subsection{Lyapunov stability}
	The following lemma gives Lyapunov's direct method for local stability; e.g., see \cite{haddad2008nonlinear}.
	\medskip
	
	\begin{lemma} \label{lemma:lyapunov_direct_method}
		{\color{red}Consider a system $\dot{x} = f(x)$ where $x \in \mathbb{R}^{n}$ is the state and $f:\mathbb{R}^{n} \to \mathbb{R}^{n}$ is a locally Lipschitz continuous function. Suppose the origin $x^{\ast} = 0$ is an equilibrium point of the system and $\mathcal{D}$ is a region containing the origin.
		If there exists a continuously differentiable function $V(x): \mathcal{D} \to \mathbb{R}$ such that
		\begin{subequations}
			\begin{align}
				&V(x) > 0, \forall x \in \mathcal{D} \setminus \{0\}, \quad V(0) = 0, \  \text{and} \label{eq:con1}\\
				&\dot{V}(x) = \frac{\partial V}{\partial x}f(x) \leq 0, \forall x \in \mathcal{D}, \label{eq:con2}
			\end{align}
		\end{subequations}
		then the origin  is locally stable in the sense of Lyapunov. If, additionally, we have 
		\begin{equation}
			\dot{V}(x) = \frac{\partial V}{\partial x}f(x) < 0, \forall x \in \mathcal{D} \setminus \{0\},
		\end{equation}
		then the origin is locally asymptotically stable. }
	\end{lemma} 
	
	\subsection{Negative Imaginary Systems}
	Consider a single-input-single-output (SISO) nonlinear system with the following state-space model:
	\begin{subequations}
		\label{sys:siso_sp}
		\begin{align}
			\dot{x} &= f(x,u), \label{eq:siso_state}\\
			y & = h(x,u), \label{eq:siso_output}
		\end{align}
	\end{subequations}
	where $x \in \mathbb{R}^{n}$ is the state, $u \in \mathbb{R}$ is the input, $y \in \mathbb{R}$ is the output, $f:\mathbb{R}^{n} \times \mathbb{R} \to \mathbb{R}^{n}$ is a Lipschitz continuous function, and $h:\mathbb{R}^{n} \times \mathbb{R} \to \mathbb{R}$ is a class C$^{1}$ function. 
	\medskip
	
	\begin{assumption}\label{apt:equilibrium}
		Without loss of generality, assume $(x^{\ast},u^{\ast}) = (0,0)$ is an equilibrium point of the system~\eqref{sys:siso_sp}; i.e., $f(0,0)\equiv  0$. Moreover, {\color{red}assume} the output at the equilibrium $(0,0)$ is $y^{\ast} = h(0,0) \equiv 0$.
	\end{assumption}
	\medskip
	
	\begin{definition}
		\label{def:siso_ni}
		The system~\eqref{sys:siso_sp} is said to be a nonlinear negative imaginary (NNI) system if there exists a positive semidefinite storage function $V:\mathbb{R}^{n} \to \mathbb{R}$ of class C$^{1}$ such that  	for all $t \geq 0$,
		\begin{align}\label{eq:general_dissipation_inequality}
			\dot{V}(x) &\leq \int_{0}^{u} \frac{\partial}{\partial x}h(x,\xi)\dot{x}d\xi.
		\end{align}
	\end{definition}
	\medskip
	
	\begin{remark}
		\label{rmk:reduced_siso_ni}
		If the output in the system~\eqref{sys:siso_sp} reduces to 
		\begin{equation}
			\label{eq:siso_output_onlyx}
			y  = h(x),
		\end{equation}
		the  inequality~\eqref{eq:general_dissipation_inequality} can be simplified to
		\begin{equation}
			\label{eq:simplified_dissipation_inequality}
			\dot{V}(x) \leq u\dot{h}(x),
		\end{equation}
		which is consistent with \cite[Definition 1]{shi2023output}.
	\end{remark}
	\section{Feedback Interconnection of NNI Systems}~\label{sec:NI_Feedback}
	Consider a SISO NNI system $H_{1}$: 
	\begin{subequations}
		\label{sys:h1_sp}
		\begin{align}
			H_{1} : \quad	\dot{x}_{1} &= f_{1}(x_{1},u_{1}), \label{eq:h1_state}\\
			y_{1} & = h_{1}(x_{1}), \label{eq:h1_output}
		\end{align}
	\end{subequations}
	where $x_{1} \in \mathbb{R}^{n_{1}}$ is the state, $u_{1} \in \mathbb{R}$ is the input, $y_{1} \in \mathbb{R}$ is the output, $f_{1}:\mathbb{R}^{n_{1}} \times \mathbb{R} \to \mathbb{R}^{n_{1}}$ is a Lipschitz continuous function, and $h_{1}:\mathbb{R}^{n_{1}} \times \mathbb{R} \to \mathbb{R}$ is a class C$^{1}$ function.

	Also consider a SISO NNI system $H_{2}$: 
	\begin{subequations}
		\label{sys:h2_sp}
		\begin{align}
			H_{2} : \quad	\dot{x}_{2} &= f_{2}(x_{2},u_{2}), \label{eq:h2_state}\\
			y_{2} & = h_{2}(x_{2},u_{2}), \label{eq:h2_output}
		\end{align}
	\end{subequations}
	where $x_{2} \in \mathbb{R}^{n_{2}}$ is the state, $u_{2} \in \mathbb{R}$ is the input, $y_{2} \in \mathbb{R}$ is the output, $f_{2}:\mathbb{R}^{n_{2}} \times \mathbb{R} \to \mathbb{R}^{n_{2}}$ is a Lipschitz continuous function, and $h_{2}:\mathbb{R}^{n_{2}} \times \mathbb{R} \to \mathbb{R}$ is a class C$^{1}$ function.
	
	
	\begin{figure}
		\centering
		\includegraphics[width=0.5\linewidth]{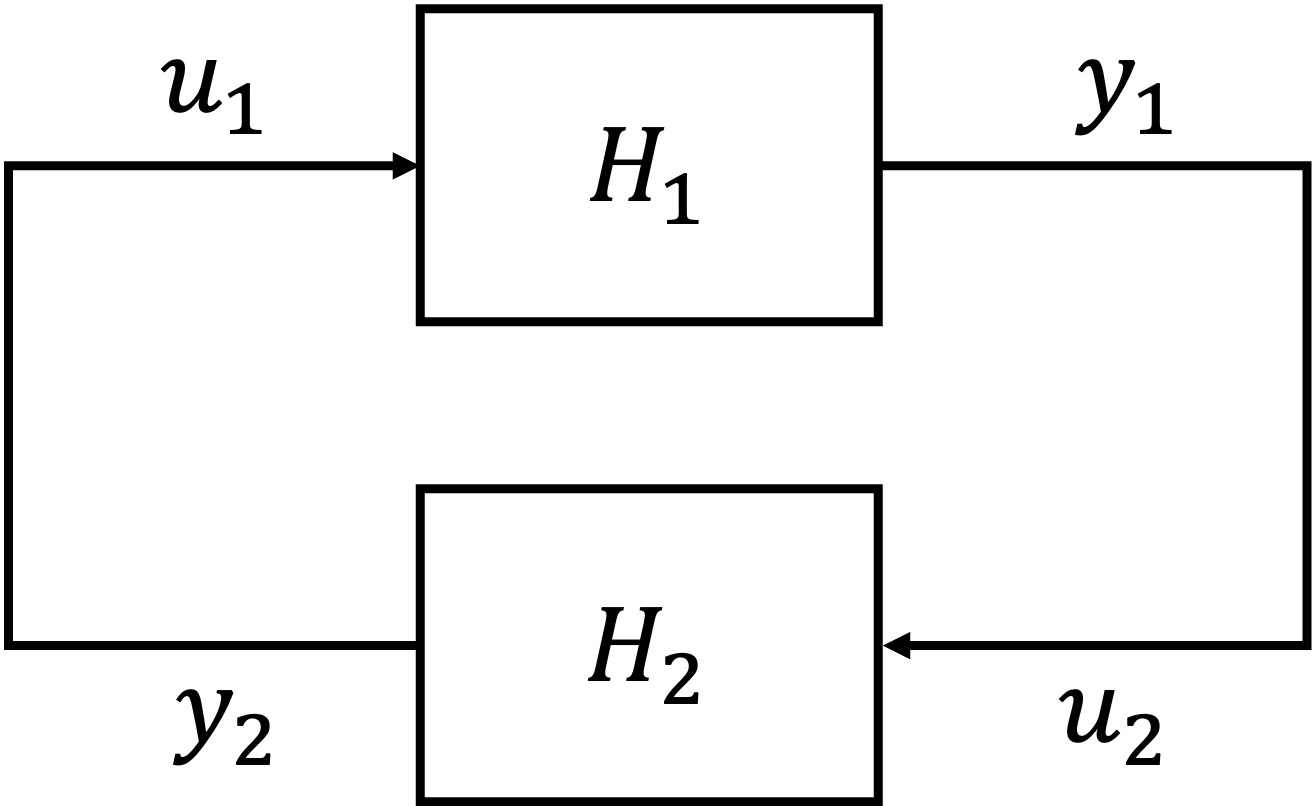}
		\caption{Feedback interconnection of  two systems $H_{1}$ and $H_{2}$.}
		\label{fig:single_loop}
	\end{figure}
	
	Further consider the feedback interconnection of  two systems $H_{1}$ and  $H_{2}$ as shown in Fig.~\ref{fig:single_loop}. The relation between the inputs  and the outputs  is described by $y_{1}  \equiv u_{2}$ and $y_{2} \equiv u_{1}$.
	
	We define {\color{red}a candidate Lyapunov} function of the feedback system $(H_{1},H_{2})$ as
	\begin{equation}
		\label{eq:fb_storage}
		W(x_{1},x_{2}) = V_{1}(x_{1}) + V_{2}(x_{2}) - \int_{0}^{h_{1}(x_{1})}h_{2}(x_{2},\xi)d\xi.
	\end{equation}
	
	\begin{assumption}
		\label{apt:local_domain}
		Suppose there exists a compact domain $\mathcal{D} \subset \mathbb{R}^{n_{1}} \times \mathbb{R}^{n_{2}}$ such that $(0,0) \in \mathcal{D}$ and $W(x_{1},x_{2}) > 0$ for $(x_{1},x_{2}) \in {\color{red}\mathcal{D} \setminus \{(0,0)\}}.$
	\end{assumption}
	\medskip
	
	\begin{remark}
		We have  constructed a novel candidate Lyapunov function~\eqref{eq:fb_storage}, fashioned in a manner reminiscent of the Lur\'e-Postnikov Lyapunov function \cite{haddad2008nonlinear}. This form points to an interesting connection between nonlinear negative imaginary systems theory and the Popov stability condition \cite{haddad2008nonlinear}; see also \cite{carrasco2017comment}.
	\end{remark}
	\medskip
	
	In the following theorem, we prove that the equilibrium $(x_{1}^{\ast},x_{2}^{\ast}) = (0,0)$ of the feedback  system $(H_{1},H_{2})$ is locally stable in the sense of Lyapunov. \medskip
	
	\begin{theorem}
		\label{thm:fb_ni}
		{\color{red}Suppose Assumption~\ref{apt:equilibrium} and Assumption~\ref{apt:local_domain} hold. Then, the equilibrium $(x_{1}^{\ast},x_{2}^{\ast}) = (0,0)$ of the feedback  system $(H_{1},H_{2})$ is locally stable in the sense of Lyapunov.}
	\end{theorem} \medskip  
	{\it Proof.} From Assumption~\ref{apt:equilibrium}, $(x_{1}^{\ast},x_{2}^{\ast}) = (0,0)$ is an equilibrium for the feedback  system $(H_{1},H_{2})$. We then want to check whether conditions \eqref{eq:con1} %
	and \eqref{eq:con2} 
	in Lemma~\ref{lemma:lyapunov_direct_method} are satisfied. First, {\color{red}using} Eq.~\eqref{eq:fb_storage} and {\color{red}the dynamics of the} NNI systems $H_{1}$ and $H_{2}$, we {\color{red}have} $W(0,0) = 0$. Also from Assumption~\ref{apt:local_domain}, we have $W(x_{1},x_{2}) > 0$ for $(x_{1},x_{2}) \in {\color{red}\mathcal{D} \setminus \{(0,0)\}}.$ Condition~\eqref{eq:con1} is thus satisfied. Second, we calculate 
	\begin{align*}
		\ &\frac{d}{dt}\int_{0}^{h_{1}(x_{1})}h_{2}(x_{2},\xi)d\xi \\
		= \ & \int_{0}^{h_{1}(x_{1})} \frac{\partial h_{2}(x_{2},\xi)}{\partial x_{2}} \dot{x}_{2} d\xi  + h_{2}(x_{2},h_{1}(x_{1}))\dot{h}_{1}(x_{1}).
	\end{align*}
	{\color{red}Using} Eqs.~\eqref{eq:general_dissipation_inequality} and~\eqref{eq:simplified_dissipation_inequality}, {\color{red}we analyze the time derivative of the candidate Lyapunov function $W(x_{1},x_{2})$:}
	\small
	\begin{align*}
		\dot{W}(x_{1},x_{2}) &=  \dot{V}_{1}(x_{1}) + \dot{V}_{2}(x_{2}) - \frac{d}{dt}\int_{0}^{h_{1}(x_{1})}h_{2}(x_{2},\xi)d\xi \\
		 &=  \dot{V}_{1}(x_{1}) + \dot{V}_{2}(x_{2}) - \int_{0}^{u_{2}} \frac{\partial h_{2}(x_{2},\xi)}{\partial x_{2}} \dot{x}_{2} d\xi  - u_{1}\dot{h}_{1}(x_{1})\\
		 &\leq  0. 
	\end{align*}
	\normalsize
	{\color{red}Thus, condition~\eqref{eq:con2} is  satisfied. Finally, Lemma~\ref{lemma:lyapunov_direct_method} is applied to conclude that the equilibrium $(x_{1}^{\ast},x_{2}^{\ast}) = (0,0)$ of the feedback system $(H_{1},H_{2})$ is locally stable in the sense of Lyapunov.} The proof is now complete. \hfill $\square$
	
	\section{Feedback Interconnection of \\ Saturated NNI Systems}\label{sec:saturated_theory}
	Consider two SISO NNI systems $H_{1}$ and $H_{3}$. The system $H_{1}$ is described by the state-space model~\eqref{sys:h1_sp}. Simultaneously, the system $H_{3}$ is characterized by:
	\begin{subequations}
		\label{sys:h3_sp}
		\begin{align}
			H_{3} : \quad \dot{x}_{3} &= f_{3}(x_{3},u_{3}), \label{eq:h3_state}\\
			y_{3} & = h_{3}(x_{3}) - Lu_{3}, \label{eq:h3_output}
		\end{align}
	\end{subequations}
	where $L \in \mathbb{R}$, $x_{3} \in \mathbb{R}^{n_{3}}$ is the state, $u_{3} \in \mathbb{R}$ is the input, $y_{3} \in \mathbb{R}$ is the output, $f_{3}:\mathbb{R}^{n_{3}} \times \mathbb{R} \to \mathbb{R}^{n_{3}}$ is a Lipschitz continuous function, and $h_{3}:\mathbb{R}^{n_{3}} \to \mathbb{R}$ is a class $C^{1}$ function.
	
	{\color{red}Suppose both systems $H_{1}$ and $H_{3}$ are NNI.}  Consequently, according to Definition~\ref{def:siso_ni}, for the system $H_{1}$, there exists a positive semidefinite storage function $V_{1}(x_{1}) \geq 0$ with $V_{1}(0) = 0$, satisfying the condition:
	\begin{align}\label{eq:h1_nni}
		&\dot{V}_{1}(x_{1}) \leq u_{1}\dot{h}_{1}(x_{1}), \quad \forall x_{1}, u_{1}.
	\end{align}
	Also, for the system $H_{3}$, there exists a positive semidefinite storage function $V_{3}(x_{3}) \geq 0$ with $V_{3}(0) = 0$, satisfying the condition:
	\begin{align}\label{eq:h3_nni}
		&\dot{V}_{3}(x_{3}) \leq	\int_{0}^{u_{3}} \frac{\partial(h_{3}(x_{3}) - L\xi)}{\partial x_{3}}\dot{x}_{3}d\xi =	u_{3}\dot{h}_{3}(x_{3}), \quad \forall x_{3},u_{3}.
	\end{align}

	\subsection{Saturated NNI Systems}
	Consider the feedback system $(H_{1}, H_{3})$. Typically, it can be interpreted as $H_{1}$ being the plant and $H_{3}$ being the controller. However, in practical implementation, {\color{red}the actuator for the controller} $H_{3}$ may be subject to saturation. As a result, we further consider an anti-windup compensator system to address potential issues that may arise from saturation. {\color{red}In particular, we consider a situation in which the controller $H_{3}$ is to replace a nominal nonlinear feedback $g(u_{3})$ via feedback linearization; see also \cite{chen2023design}. Then in the presence of actuator saturation, we consider the controller}
	\begin{subequations}
		\label{sys:tilde_h3_sp}
		\begin{align}
			\hat{H}_{3} : \ \dot{\hat{x}}_{3} &= 
			\begin{cases}
				f_{3}(\hat{x}_{3},\hat{u}_{3}), & |h_{3}(\hat{x}_{3}) - L\hat{u}_{3}- g(\hat{u}_{3})| < b,\\
				-\alpha \frac{\partial V_{3}(\hat{x}_{3})^{\top}}{\partial \hat{x}_{3}}, & |h_{3}(\hat{x}_{3}) - L\hat{u}_{3}- g(\hat{u}_{3})| \geq b,
			\end{cases} \label{eq:tilde_h3_state}\\
			\hat{y}_{3} & = \hat{h}_{3}(\hat{x}_{3},\hat{u}_{3}) \notag\\
			&= g(\hat{u}_{3}) + \sat_{b}(h_{3}(\hat{x}_{3}) - L\hat{u}_{3}- g(\hat{u}_{3})), \label{eq:tilde_h3_y2}
		\end{align}
	\end{subequations}
	where $\alpha >0$ is a design choice, $b \geq 0$ is the saturation parameter, and $g:\mathbb{R} \to \mathbb{R}$ is a class C$^{1}$ function such that $|L\hat{u}_{3}+ g(\hat{u}_{3})| < b$ for $\hat{u}_{3} \in \mathcal{D}_{\hat{u}_{3}}$. The domain $\mathcal{D}_{\hat{u}_{3}}$ represents the domain of interest for the output of the system $H_{1}$. For example, in power systems, $\mathcal{D}_{\hat{u}_{3}}$ is the domain of attraction for  rotor angles.  Here, $\sat_{b}(w)$ represents the saturation function 
	$$
	\sat_{b}(w) = 	
	\begin{cases}
		-b, & w \leq -b,\\
		w, & -b < w < b,\\
		b,& w\geq  b,
	\end{cases}
	$$
	where the saturation variable $w$ is defined to be  $$w = h_{3}(\hat{x}_{3}) - L\hat{u}_{3}- g(\hat{u}_{3}).$$
	
	\begin{remark}
				When $g(\hat{u}_{3}) \equiv 0$,  the output ~\eqref{eq:tilde_h3_y2} in the system $\hat{H}_{3}$
				reduces to $\hat{y}_{3} = \sat_{b}(h_{3}(\hat{x}_{3}) - L\hat{u}_{3}) = \sat_{b}(y_{3})$, characterizing actuator saturation of the output~\eqref{eq:h3_output} in the system $H_{3}$. To ensure the applicability of our theory to power systems, we propose the more general form \eqref{eq:tilde_h3_y2}  for the controller $\hat{H}_{3}$ to replace a nominal nonlinear feedback $g(\hat{u}_{3})$ via feedback linearization in which $g(\hat{u}_{3})$ represents the sine function applied to the rotor angle as detailed in Section~\ref{sec:local_storage}.
	\end{remark}
	\medskip

	\begin{remark}
		As shown in the following Theorem~\ref{thm:tilde_h3_nni} and Remark~\ref{rmk:advantage_compensator}, the advantages of such a control design for the controller $\hat{H}_{3}$ are two-fold: (1) it ensures that the controller $\hat{H}_{3}$ maintains NNI property, (2) and it ensures that the actuator {\color{red}does not stay at} the saturation boundary as time goes to infinity. 
	\end{remark}
	\medskip
	
	\begin{theorem}
		\label{thm:tilde_h3_nni}
		The  controller $\hat{H}_{3}$ is  NNI. 
	\end{theorem} \medskip 
	{\it Proof.} We take $V_{3}(\hat{x}_{3})$ as the storage function of the system $\hat{H}_{3}$ and analyze its derivative with respect to time:
	\begin{equation*}
		\dot{V}_{3}(\hat{x}_{3}) = \begin{cases}
			\frac{\partial V_{3}(\hat{x}_{3})}{\partial \hat{x}_{3}} f_{3}(\hat{x}_{3},\hat{u}_{3}), & |w| < b,\\
			-\alpha\| \frac{\partial V_{3}(\hat{x}_{3})}{\partial \hat{x}_{3}}\|^{2}, & |w| \geq b.
		\end{cases} 
	\end{equation*}
	We also calculate
		\begin{align}\label{eq:dot_V_3}
			&\int_{0}^{\hat{u}_{3}} \frac{\partial \hat{h}_{3}(\hat{x}_{3},\xi)}{\partial \hat{x}_{3}}\dot{\hat{x}}_{3}d\xi
			=  \begin{cases}
				\hat{u}_{3}\dot{h}_{3}(\hat{x}_{3}), & |w| < b,\\
				0, & |w| \geq b.
			\end{cases} 
		\end{align}
	Under both cases,  $\dot{V}_{3}(\hat{x}_{3}) \leq \int_{0}^{\hat{u}_{3}} \frac{\partial \hat{h}_{3}(\hat{x}_{3},\xi)}{\partial \hat{x}_{3}}\dot{\hat{x}}_{3}d\xi$ holds. Hence, its follows from Definition~\ref{def:siso_ni} that the  system $\hat{H}_{3}$ is NNI. The proof is now complete. \hfill $\square$
	\medskip
	
	\begin{remark}\label{rmk:advantage_compensator}
		It is noted that  {\color{red}the definition of the system $\hat{H}_{3}$ ensures that the actuator does not stay at the saturation boundary as time goes to infinity}. For all 
		$$ \hat{x}_{3} \in \mathcal{D}_{sat} = \{\hat{x}_{3} : |w| \geq  b \},$$
		we have $\dot{V}_{3}(\hat{x}_{3}) = -\alpha\| \frac{\partial V_{3}(\hat{x}_{3})}{\partial \hat{x}_{3}}\|^{2} < 0$, which implies that $V_{3}(\hat{x}_{3})$ will keep decreasing when $\hat{x}_{3} \in \mathcal{D}_{sat}$. Since $V_{3}(0) = 0$,  $h_{3}(0)$, $|L\hat{u}_{3}+ g(\hat{u}_{3})| < b$, and the functions $V_{3}(\cdot)$ and $h_{3}(\cdot)$ are continuous, it follows that $\hat{x}_{3}$ will eventually leave the saturation region $\mathcal{D}_{sat}$. 
	\end{remark}

	\subsection{Lyapunov Stability of the Feedback Interconnection}
	 Consider the feedback system $(H_{1},\hat{H}_{3})$. 
	 The input-output relation is described by 
	\begin{subequations}
		\begin{align}
			&u_{1} \equiv \hat{y}_{3} = g(\hat{u}_{3}) + \sat_{b}(w),\\
			&\hat{u}_{3} \equiv y_{1} = h_{1}(x_{1}).
		\end{align}
	\end{subequations}
	The system $H_{1}$ {\color{red}is assumed to be NNI} and the system $\hat{H}_{3}$ is proved to be NNI in Theorem~\ref{thm:tilde_h3_nni}.  Consequently, according to Definition~\ref{def:siso_ni}, for the system $H_{1}$, a positive semidefinite storage function $V_{1}(x_{1}) \geq 0$ with $V_{1}(0) = 0$ exists, satisfying the condition:
		\begin{align}\label{eq:h1_h3_nni}
			& \dot{V}_{1}(x_{1}) \leq 
			\begin{cases}
				(g(\hat{u}_{3}) - b)\dot{h}_{1}(x_{1}),&w \leq -b,\\
				(h_{3}(\hat{x}_{3}) - L\hat{u}_{3})\dot{h}_{1}(x_{1}), & |w| < b,\\
				(g(\hat{u}_{3}) + b)\dot{h}_{1}(x_{1}),&w \geq b,
			\end{cases} 
		\end{align}
	{\color{red}for all $ x_{1}, \hat{x}_{3}, \hat{u}_{3}$.}
	For the system $H_{3}$, a positive semidefinite storage function $V_{3}(x_{3}) \geq 0$ with $V_{3}(0) = 0$ exists, satisfying the condition:
		\begin{align}\label{eq:h3_h1_nni}
			\dot{V}_{3}(\hat{x}_{3}) \leq			
			 \begin{cases}
					h_{1}(x_{1})\dot{h}_{3}(\hat{x}_{3}), & |w| < b,\\
					0, & |w| \geq b,
				\end{cases} 
		\end{align}
	{\color{red}for all $x_{1}, \hat{x}_{3}.$}

	We further define  a function  $F$  to represent  $\int_{0}^{h_{1}(x_{1})}\hat{h}_{3}(\hat{x}_{3},\xi)d\xi$:
	\begin{equation*}\label{eq:F}
		F = 
		\begin{cases}
			\int_{0}^{h_{1}(x_{1})} \big(g(\xi) -b\big) d\xi, & w \leq - b,\\
			\int_{0}^{h_{1}(x_{1})} \big(h_{3}(\hat{x}_{3}) - L\xi\big)d\xi, & -b < w < b,\\
			\int_{0}^{h_{1}(x_{1})} \big(g(\xi) +b\big) d\xi,  & w \geq b, 
		\end{cases}
	\end{equation*}
	and calculate its derivative with respect to time as
	\small
	\begin{equation}\label{eq:dot_F}
		\dot{F} = 
		\begin{cases}
			\big(g(h_{1}(x_{1})) -b\big)\dot{h}_{1}(x_{1}), & w \leq - b,\\
			\big(h_{3}(\hat{x}_{3}) - Lh_{1}(x_{1})\big)\dot{h}_{1}(x_{1})  + h_{1}(x_{1})\dot{h}_{3}(\hat{x}_{3}), & |w| < b,\\
			\big(g(h_{1}(x_{1})) +b\big)\dot{h}_{1}(x_{1}),  & w \geq b.
		\end{cases}
	\end{equation}
	\normalsize
	As in Eq.~\eqref{eq:fb_storage}, {\color{red}a candidate Lyapunov function} for the feedback system $(H_{1},\hat{H}_{3})$ is described by 
	\begin{align*}
		\label{eq:fb_storage_hat_h3}
		\hat{W}(x_{1},\hat{x}_{3}) 
		= \ &V_{1}(x_{1}) + V_{3}(\hat{x}_{3}) - \int_{0}^{h_{1}(x_{1})}\hat{h}_{3}(\hat{x}_{3},\xi)d\xi \\
		= \ &  V_{1}(x_{1}) + V_{3}(\hat{x}_{3}) - F.
	\end{align*}

	\begin{assumption}
		\label{apt:local_domain_hat_h3}
		Suppose there exists a compact domain $\mathcal{D} \subset \mathbb{R}^{n_{1}} \times \mathbb{R}^{n_{3}}$ such that $(0,0) \in \mathcal{D}$ and $\hat{W}(x_{1},\hat{x}_{3}) > 0$ for $(x_{1},\hat{x}_{3}) \in {\color{red}\mathcal{D} \setminus \{(0,0)\}}.$
	\end{assumption}
	\medskip
	
	In the following theorem, we show that the equilibrium $(x_{1}^{\ast},\hat{x}_{3}^{\ast}) = (0,0)$ of the feedback  system $(H_{1},\hat{H}_{3})$ is locally stable  in the sense of Lyapunov. \medskip
	
	\begin{theorem}
		\label{thm:fb_ni_hat_h3}
		{\color{red}Suppose Assumption~\ref{apt:equilibrium} and Assumption~\ref{apt:local_domain} hold. Then, the equilibrium $(x_{1}^{\ast},\hat{x}_{3}^{\ast}) = (0,0)$ of the feedback  system $(H_{1},\hat{H}_{3})$ is locally stable  in the sense of Lyapunov.}
	\end{theorem} \medskip 
	{\it Proof.} Indeed, $(x_{1}^{\ast},\hat{x}_{3}^{\ast}) = (0,0)$ is an equilibrium point for the feedback  system $(H_{1},\hat{H}_{3})$. 	We then want to check whether conditions \eqref{eq:con1} and \eqref{eq:con2}  in Lemma~\ref{lemma:lyapunov_direct_method} are satisfied. First, we can calculate $\hat{W}(0,0) = 0$. Also from Assumption~\ref{apt:local_domain_hat_h3}, we have $\hat{W}(x_{1},\hat{x}_{3}) > 0$ for $(x_{1},\hat{x}_{3}) \in {\color{red}\mathcal{D} \setminus \{(0,0)\}}.$ Condition~\eqref{eq:con1} is thus satisfied. Second, according to Eqs.~\eqref{eq:h1_h3_nni},~\eqref{eq:h3_h1_nni}, and \eqref{eq:dot_F}, {\color{red}we calculate the time derivative of the candidate Lyapunov function as}
	\begin{align*}
		\dot{\hat{W}}(x_{1},\hat{x}_{3}) =
	\dot{V}_{1}(x_{1})   + \dot{V}_{3}(\hat{x}_{3}) - \dot{F}\leq 0.
	\end{align*}
	Hence, 
	condition~\eqref{eq:con2} is  satisfied. Finally, Lemma~\ref{lemma:lyapunov_direct_method} is applied to conclude that the equilibrium $(x_{1}^{\ast},\hat{x}_{3}^{\ast}) = (0,0)$ of the feedback  system $(H_{1},\hat{H}_{3})$ is locally stable  in the sense of Lyapunov.  The proof is now complete. \hfill $\square$

	\section{Single-Machine-Infinite-Bus Power Systems}~\label{sec:SMIB}
\begin{figure}[tb]
		\centering
		\includegraphics[width=0.37\textwidth]{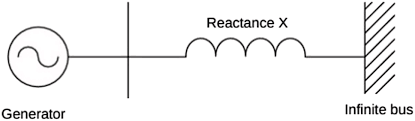}
		\caption{A single-machine-infinite-bus power system.}
		\label{fig:smib}
	\end{figure}

	Consider a single-machine-infinite-bus  (SMIB) power system. As illustrated in Fig.~\ref{fig:smib}, a generator is connected to an infinite bus representing the bulk power grid.  For the infinite bus, the voltage magnitude $|V_{s}|$ is assumed to be constant and the voltage phase is assumed to be zero. For the generator bus, we denote the rotor angle in the stationary reference frame by $\theta$, the rotor speed by $\omega$, and the nominal frequency by $\omega^{0}$.  Then, the voltage at time $t$ is written as a cosine function multiplied by its  voltage magnitude $|V_{g}|$: 
		\begin{align*}
			|V_{g}|\cos(\theta(t)) &= |V_{g}|\cos(\omega^{0}t + \delta(t)) \\
			&= |V_{g}|\cos(\omega^{0}t + \bar{\delta} + \tilde{\delta}(t)),
		\end{align*}
	where $\delta(t)$ represents the voltage phase angle, $\bar{\delta}$ represents the voltage phase angle at steady state, and $\tilde{\delta}(t) = \delta(t) - \bar{\delta}$ represents the  voltage phase angle deviation. The frequency on the generator bus is defined as 
	\begin{equation}\label{eq:frequency_def}
		\omega = \dot{\theta} =\omega^{0} + \dot{\tilde{\delta}}.
	\end{equation}

	Based on the generator mechanical model \cite{glover2012power}, the swing equation of the generator bus  is described by  
	\begin{equation}\label{eq:smib_d}
		M\ddot{\delta}= P^{M}  - P^{E} - D\dot{\delta},
	\end{equation}
	where $M = \frac{2H}{\omega^{0}}  >  0$ represents inertia on the generator bus with a lumped parameter $H >0 $, and $D \geq 0$ is the damping coefficient. Here, $P^{M}$ represents the mechanical power injection on the generator bus, and $P^{E} $ represents the electric power output of the generator bus. The electric power output  $P^{E} $ of the generator bus  is given by 
	\begin{equation}\label{eq:electric_output}
		P^{E}= \frac{|V_{g}||V_{s}|}{X}\sin \delta = P_{\max}^{E}\sin \delta, 
	\end{equation}
	where  $X$ is the total reactance, and $P_{\max}^{E}$ is the maximum power transfer between the generator bus and the infinite bus, and $|V_{g}|$ and $|V_{s}|$ are at the nominal voltage.
	\medskip

	
	{\noindent \bf Fault Setting.} In practice, sudden changes in electrical or mechanical power can arise. Prior to a fault occurrence, the SMIB system~\eqref{eq:smib_d} is {\color{red}assumed to be} at a stable equilibrium given by
	$	(\bar{\delta}, \bar{P}^{M},\bar{P}^{E}_{\max}),$
	where it satisfies the relation $\bar{P}^{M} = \bar{P}^{E}_{\max}\sin \bar{\delta}$. 
	In the aftermath of a fault, the generator bus frequency deviates from its nominal value $\omega^{0}$. The phase angle deviates from its steady-state value $\bar{\delta}$ and may reach a new steady-state value $\delta_{0}$ caused by sudden changes in mechanical or electrical power.  In this paper, we specifically focus on the post-fault transient stability, during which the mechanical power and maximum power transfer return to their pre-fault values $(\bar{P}^{M}, \bar{P}^{E}_{\max})$.

	\subsection{Equal Area Criterion}

	In what follows, we present the equal area criterion, a direct method for determining post-fault transient stability for SMIB systems without damping \cite{glover2012power}.
	
\begin{figure}[tb]
		\centering
		\includegraphics[width=0.8\linewidth]{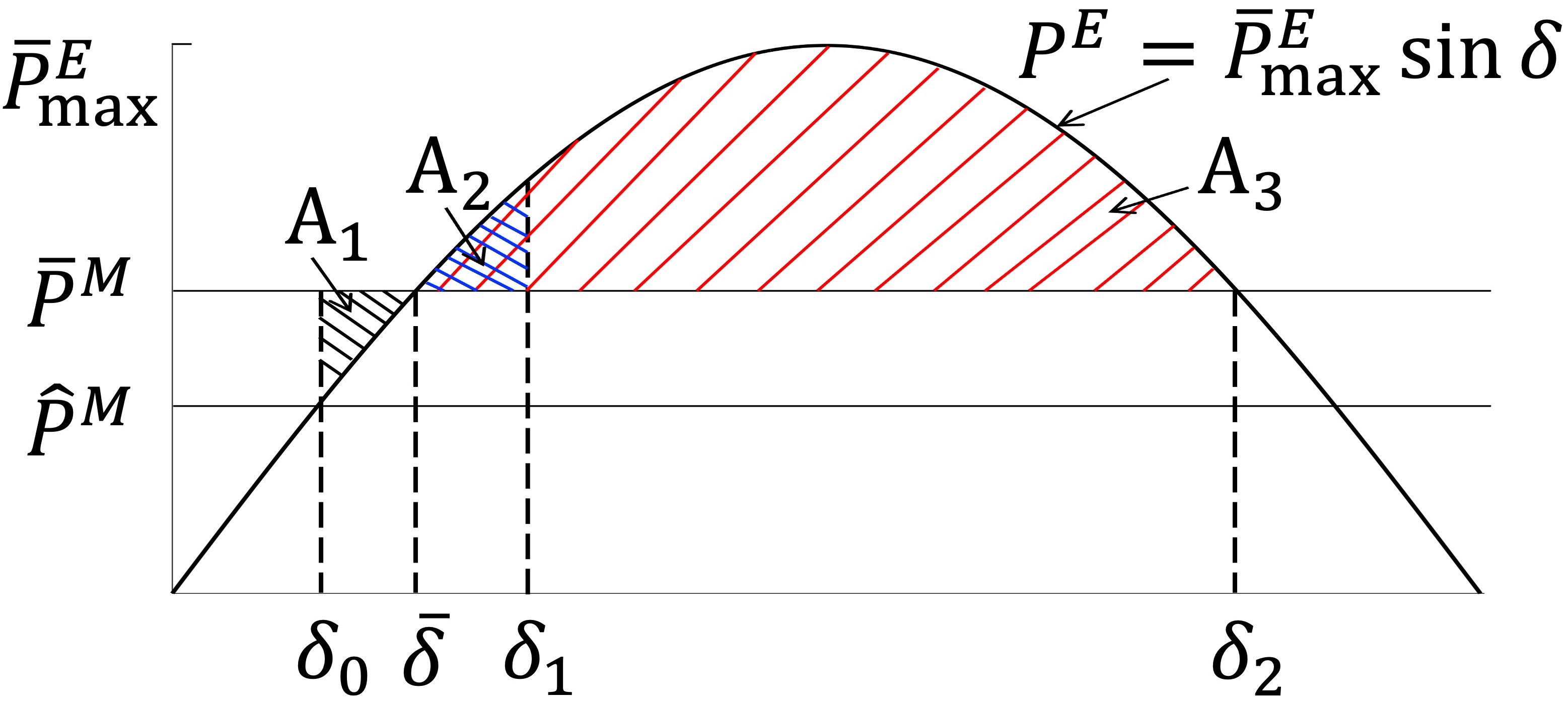}
		\caption{Mechanical power and electrical power versus angle.}
		\label{fig:EAC}
	\end{figure}

	Consider a scenario where a fault occurs leading to a change in mechanical power while the maximum power transfer remains unaffected. Denote the mechanical power during the fault as $\hat{P}^{M}$. Assume that the phase angle stabilizes at a steady-state value $\delta_{0}$ shortly before the fault is resolved. Subsequently, after the fault is cleared, the mechanical power returns to its pre-fault value $\bar{P}^{M}$. We illustrate the mechanical power and electrical power versus power angle in Fig.~\ref{fig:EAC}. 
	
	Starting from the initial angle $\delta_{0}$, the angle accelerates and $\delta$ increases. When $\delta$ reaches $\bar{\delta}$,  $\dot{\delta}$ is still positive and $\delta$ continues to increase with a decreasing speed. Eventually, $\delta$ reaches a maximum value $\delta_{1}$ and then swings back toward $\bar{\delta}$. 
	
	 The equal area criterion states that the accelerating area must be equal to the decelerating area, i.e., $\text{A}_{1}=\text{A}_{2}$. It is noted that  $\delta_{1}$ cannot exceed $\delta_{2} = \pi - \bar{\delta}$. Otherwise, the mechanical power injection would exceed the electric power output and the generator bus would accelerate again, leading to a further increase in $\delta$ and loss of stability.
	In other words, the system can maintain stability around $ \bar{\delta}$, if $\text{A}_{1} - \text{A}_{3} < 0$, equivalently,
	\begin{equation}\label{eq:eac}
		(\pi - \bar{\delta} - \delta_{0})\sin\bar{\delta} - \cos\bar{\delta}-\cos\delta_{0} < 0.
	\end{equation}

	\subsection{Lyapunov Stability of SMIB Systems}
	In what follows, we show that the SMIB system is locally stable in the sense of Lyapunov.
	\medskip
	
	The swing equation of the SMIB system can be rewritten in terms of the angle deviation $\tilde{\delta}$ specifically:
	\begin{equation}\label{eq:smib_deviation_damp}
		M\ddot{\tilde{\delta}} + D\dot{\tilde{\delta}} = \bar{P}^{E}_{\max}\sin\bar{\delta} - \bar{P}_{\max}^{E}\sin(\tilde{\delta} + \bar{\delta}) ,
	\end{equation}
	with initial angle deviation $\tilde{\delta}_{0} = \delta_{0} - \bar{\delta} \neq 0$. By reformulating the system~\eqref{eq:smib_deviation_damp} into the feedback interconnection of two NNI systems,  we can prove that the SMIB system  is locally stable in the sense of Lyapunov.
	\medskip
	
	Define $x_{1} = [\dot{\tilde{\delta}},  \tilde{\delta} ]^{\top} \in \mathbb{R}^{2}$, $u_{1} \in \mathbb{R}$, $x_{2} \equiv  0$, and $u_{2} \in \mathbb{R}$. The system \eqref{eq:smib_deviation_damp}  can be represented by the interconnection of two systems:
	\begin{subequations}
		\label{sys:smib_reformulate_with_damp}
		\begin{align}
			H_{1} : \quad	\dot{x}_{1} &= Ax_{1} + Bu_{1} \label{eq:damp_smib_h1_x}\\
			y_{1} & = Cx_{1} \label{eq:damp_smib_h1_y} \\
			H_{2} :  \quad y_{2} & = \bar{P}^{E}_{\max}\sin\bar{\delta}  -  P_{\max}^{E}\sin(u_{2} + \bar{\delta}) \label{eq:damp_smib_h2}
		\end{align}
	\end{subequations}
	where $H_{1}$  has system matrices 
	$$
	A = \begin{bmatrix}
		\frac{-D}{M} & 0 \\
		1 & 0 
	\end{bmatrix}, \ B = \begin{bmatrix}
		\frac{1}{M} \\
		0
	\end{bmatrix},  \ C = \begin{bmatrix}
		0 & 1
	\end{bmatrix},
	$$
	and the feedback system $(H_{1},H_{2})$ {\color{red}is such that} $u_{1} \equiv y_{2}$ and  $u_{2} \equiv y_{1}$. 
	\medskip
	
	In the subsequent theorem, establishing the NI nature of both systems lays the ground for proving the local stability of the SMIB system in the sense of Lyapunov.
	\medskip
	
	\begin{theorem}
		{\color{red}The equilibrium $(\dot{\tilde{\delta}}^{\ast},\tilde{\delta}^{\ast}) = (0,0)$ of the system \eqref{eq:smib_deviation_damp} is locally stable  in the sense of Lyapunov.}
	\end{theorem} \medskip 
	{\it Proof.} From the system \eqref{eq:smib_deviation_damp}, $(\dot{\tilde{\delta}}^{\ast},\tilde{\delta}^{\ast}) = (0,0)$ is an equilibrium. The system \eqref{eq:smib_deviation_damp} can be represented by the interconnection of two systems in ~\eqref{sys:smib_reformulate_with_damp}. As long as we can show both two system $H_{1}$ and $H_{2}$ are NI and verify Assumption~\ref{apt:local_domain}, Theorem~\ref{thm:fb_ni} can be applied to prove that $(\dot{\tilde{\delta}}^{\ast},\tilde{\delta}^{\ast}) = (0,0)$ is a locally stable equilibrium in the sense of Lyapunov.
	
	We first show that both systems $H_{1}$ and  $H_{2}$ in \eqref{sys:smib_reformulate_with_damp} are NI. For system $H_{1}$, we choose its storage function to be $$V_{1}(x_{1}) = x_{1}^{\top}Px_{1} \geq 0$$ 
	with  $P = \begin{bmatrix}
		\frac{M}{2} & 0\\ 0 & 0
	\end{bmatrix}.$ We then calculate 
	\begin{align*}
		\dot{V}_{1}(x_{1})
		= \ &  -Dx_{1}^{\top}\begin{bmatrix}
			1 & 0\\
			0 & 0 
		\end{bmatrix}x_{1}+ u_{1}\dot{y}_{1} \leq  u_{1}\dot{y}_{1}.
	\end{align*}
	Hence, according to Definition~\ref{def:siso_ni}, the system $H_{1}$ is NI. 
	For the system $H_{2}$, there are no associated dynamics and the state $x_{2}$ stays at zero; i.e., $x_{2}\equiv 0$. The storage function of the system $H_{2}$ can be considered as zero. Hence, according to Definition~\ref{def:siso_ni}, the system $H_{2}$ is also NI.
	
	We next verify Assumption~\ref{apt:local_domain}. 		Consider a  domain $\mathcal{D}=\{(\dot{\tilde{\delta}},\tilde{\delta}) |\cos\bar{\delta} > \cos(\tilde{\delta}+\bar{\delta}) + \tilde{\delta}\sin\bar{\delta}\} \cup (0,0)$. For ${\color{red}\mathcal{D} \setminus \{(0,0)\}}$, we analyze the {\color{red} candidate Lyapunov} function of the feedback system $(H_{1},H_{2})$ in \eqref{sys:smib_reformulate_with_damp}:
	\begin{align*}
		W(x_{1}) 
		=\ &  x_{1}^{\top}Px_{1} - \bar{P}^{E}_{\max}\int_{0}^{\tilde{\delta}}\Big(\sin\bar{\delta} - \sin(\xi+ \bar{\delta})\Big)d\xi\\
		=\ & x_{1}^{\top}Px_{1} + \bar{P}^{E}_{\max} \big(\cos\bar{\delta} - \tilde{\delta}\sin\bar{\delta}- \cos(\tilde{\delta}+\bar{\delta}) \big)\\
		>&0.
	\end{align*}
	Theorem~\ref{thm:fb_ni} is thus applied to assert that {\color{red}the equilibrium $(x_{1}^{\ast},x_{2}^{\ast}) = (0,0,0)$ of the feedback system $(H_{1}, H_{2})$  is locally stable in the sense of Lyapunov}. It follows that the equilibrium $(\dot{\tilde{\delta}}^{\ast},\tilde{\delta}^{\ast}) = (0,0)$ of the system \eqref{eq:smib_deviation_damp} is locally stable  in the sense of Lyapunov. 
	The proof is now complete. \hfill$\square$
	
	When $D > 0$, we analyze the time derivative of the candidate Lyapunov function:
	\begin{align*}
		\dot{W}(x_{1}) 
		= \ &\dot{V}_{1}(x_{1})  - h_{2}(h_{1}(x_{1}))\dot{h}_{1}(x_{1})\\
		= & -Dx_{1}^{\top}\begin{bmatrix}
			1 & 0\\
			0 & 0 
		\end{bmatrix}x_{1}+ u_{1}\dot{y}_{1} - u_{1}\dot{y}_{1}\\
		= & -  D\dot{\tilde{\delta}}^{2}.
	\end{align*}
	From this, it follows that $\dot{W}(x_{1}) = 0$ is only possible when $\dot{\tilde{\delta}} = 0$. Hence, $\dot{W}(x_{1})$ can remain zero only if  $\dot{\tilde{\delta}}$ remains zero, i.e., $\dot{W}(x_{1}) \equiv 0 \Rightarrow  \dot{\tilde{\delta}} \equiv 0.$ According to the feedback dynamics \eqref{sys:smib_reformulate_with_damp}, we have $\dot{\tilde{\delta}} \equiv 0 \Rightarrow u_{1} \equiv 0 \Rightarrow y_{2} \equiv 0 \Rightarrow \tilde{\delta} \equiv 0 \Rightarrow x_{1} \equiv [0 \ 0]^{\top}.$ 
	Thus, we can conclude that $\dot{W}(x_{1})$ cannot remain zero unless $x_{1} = [ 0 , 0]^{\top}$. Therefore, according to LaSalle's invariance principle, {\color{red}the equilibrium $(x_{1}^{\ast},x_{2}^{\ast}) = (0,0,0)$ of the feedback system $(H_{1},H_{2})$ is locally asymptotically stable.} This implies that the equilibrium $(\dot{\tilde{\delta}}^{\ast},\tilde{\delta}^{\ast}) = (0,0)$ of the system \eqref{eq:smib_deviation_damp} with damping  is locally asymptotically stable.
	
	\medskip
	
	\begin{remark}\label{rmk:SMIB_stability}
		{\color{red}In the absence of the damping $(D = 0)$, the equilibrium $(\dot{\tilde{\delta}}^{\ast},\tilde{\delta}^{\ast}) = (0,0)$ of the system~\eqref{eq:smib_deviation_damp} is locally stable in the sense of Lyapunov. In the presence of the damping $(D > 0)$, the equilibrium $(\dot{\tilde{\delta}}^{\ast},\tilde{\delta}^{\ast})= (0,0)$ of the system~\eqref{eq:smib_deviation_damp} is locally asymptotically stable.}
	\end{remark}
	
	\subsection{Invariant Set}
	Since the system \eqref{eq:smib_deviation_damp} may have several equilibria,  stability is not sufficient to guarantee that the system converges to the equilibrium at $(0,0)$. We need to find {\color{red}an} invariant set so that any initial state starting from the region will remain in {\color{red}this} region. {\color{red}An} invariant subset of $\mathcal D$ is considered as
	\begin{equation}\label{eq:level set}
		\Omega_c = \{x_1\in \mathbb R^2|W(x_1)\leq c\}.
	\end{equation}
	We find the constant {\color{red}$c >0$} as follows. Consider the form of the Lyapunov function
	\begin{equation*}
		W(\tilde \delta,\dot {\tilde \delta}) = \frac{M}{2}\dot {\tilde \delta}^2- \bar{P}^{E}_{\max}\tilde{\delta}\sin\bar{\delta} + \bar{P}^{E}_{\max}\cos\bar{\delta}- \bar{P}^{E}_{\max}\cos(\tilde{\delta}+\bar{\delta}).
	\end{equation*}
	Since the term $M\dot {\tilde \delta}^2$ increase monotonically with $|\dot{\tilde{\delta}}|$, then we only analyze the function
	\begin{equation*}
		\Gamma (\tilde \delta)  = - \bar{P}^{E}_{\max}\tilde{\delta}\sin\bar{\delta} + \bar{P}^{E}_{\max}\cos\bar{\delta}- \bar{P}^{E}_{\max}\cos(\tilde{\delta}+\bar{\delta}).
	\end{equation*}
	We take the derivative of $\Gamma (\tilde \delta)$ with respect to $\tilde \delta$.
	\begin{equation*}
		\frac{d}{d\tilde \delta} \Gamma (\tilde \delta) = - \bar{P}^{E}_{\max}\sin\bar{\delta}+\bar{P}^{E}_{\max}\sin(\tilde{\delta}+\bar{\delta}).
	\end{equation*}
	Letting $\frac{d}{d\tilde \delta} \Gamma (\tilde \delta)=0$, we have that
	$\sin(\tilde{\delta}+\bar{\delta})=\sin\bar{\delta}$. That is $\tilde \delta = 2k\pi$ or $\tilde \delta = -2\bar \delta + (2k+1)\pi$. We only focus on a neighborhood of the origin. 
	The function $\Gamma (\tilde \delta)$ has two local maxima near the origin. They are $\Gamma(-\pi-2\bar \delta) = \bar P_{max}^E(2\cos(\bar \delta)+(\pi-2\bar \delta)\sin(\bar \delta))$ and $\Gamma(\pi-2\bar \delta) = \bar P_{max}^E(2\cos(\bar \delta)-(\pi-2\bar \delta)\sin(\bar \delta))$. Since $\bar \delta \in (0,\frac{\pi}{2})$, then $\Gamma(\pi-2\bar \delta)<\Gamma(-\pi-2\bar \delta).$ Hence, we have
	\begin{equation}\label{eq:c}
		0<c<\Gamma(\pi-2\bar \delta)=\bar P_{max}^E(2\cos(\bar \delta)-(\pi-2\bar \delta)\sin(\bar \delta)).
	\end{equation}
	Note that the set (\ref{eq:level set}) with (\ref{eq:c}) may {\color{red}consist several disconnected sets.} The one that contains the origin is a subset of $\mathcal D$. To restrict the states to this set, we give an extra upper bound {\color{red}on} the state $\tilde \delta$. Denoting $x_{1}^{[2]}:=\tilde \delta$, we have
	\begin{equation}\label{eq:Region of Attraction}
		\Omega = \{x_1\in \mathbb R^2|W(x_1)\leq c \textnormal{ and } |x_{1}^{[2]}|\leq \pi-2\bar \delta\},
	\end{equation}
	where $c$ is given in (\ref{eq:c}). Therefore, the set (\ref{eq:Region of Attraction}) is the {\color{red}required} compact invariant set.
	
	\medskip
	
	{\noindent \bf Comparison with Equal Area Criterion.} Assuming the SMIB system starts at  initial state $(\dot{\tilde{\delta}}, \tilde{\delta}) = (\delta_{0},0)$. According to Eq.~\eqref{eq:Region of Attraction}, the invariant set for $\tilde{\delta}_{0}$ is described by 
	\[ \{\tilde{\delta} : (\pi - 2\bar{\delta} - \tilde{\delta})\sin\bar{\delta} - \cos\bar{\delta} - \cos(\bar{\delta} + \tilde{\delta}) < 0\}. \]
	By shifting $\tilde{\delta}_{0}$ to $\delta_{0} = \bar{\delta} + \tilde{\delta}_{0}$, the invariant set for $\delta_{0}$ becomes
	\begin{equation}\label{eq:delta_invariant_set}
		\{\delta : (\pi - \bar{\delta} - \delta)\sin\bar{\delta} - \cos\bar{\delta} - \cos(\delta) < 0\},
	\end{equation}
	which aligns with the stability condition~\eqref{eq:eac} derived from the equal area criterion. Notably, the stability condition~\eqref{eq:eac} derived from the equal area criterion represents a specific case within our analysis of the invariant set~\eqref{eq:Region of Attraction}, assuming $\dot{\delta}_{0} = 0$. Thus, our exploration of the invariant set serves as a theoretical foundation for the equal area criterion from  perspective of nonlinear systems theory.

	
	\section{SMIB Systems with  Battery-based Control}~\label{sec:local_storage}
	In what follows, our focus lies in equipping the generator bus  with a large-scale battery and providing a systematic method to design  angle based feedback control based on the use of the battery-based actuator to enhance the transient stability of SMIB systems.
	\medskip
	
	We consider a battery equipped at the generator bus. Hence, modifying Eq.~\eqref{eq:smib_d}, the swing equation is revised as 
	\begin{equation}\label{eq:storage_smib_d}
		M\ddot{\delta}= P^{M} + P^{ST}  - P^{E} - D\dot{\delta},
	\end{equation}
	where $P^{ ST}$ represents the power output of the local storage device at the generator bus. {\color{red}It is noted that the battery has a finite maximum power output. }  
	\medskip
	
	Prior to a fault occurrence, the system~\eqref{eq:storage_smib_d} is at a stable equilibrium 
$		(\bar{\delta}, \bar{P}^{M}, \bar{P}^{ST}, \bar{P}^{E}_{\max}),$
	where
	$	\bar{P}^{M} + \bar{P}^{ST}- \bar{P}^{E}_{\max}\sin \bar{\delta} = 0.$ 
	In the aftermath of a fault, the generator bus frequency deviates from its nominal value $\omega^{0}$. The phase angle deviates from its steady-state value $\bar{\delta}$ and may reach a new stable-state value $\delta_{0}$ caused by sudden changes in mechanical or electrical power. In what follows, we investigate the post-fault transients, during which mechanical power and maximum power transfer are  back to pre-fault values $(\bar{P}^{M},\bar{P}^{E}_{\max})$. We aim to design battery-based angle feedback control to stabilize the SMIB system even when equal area criterion~\eqref{eq:eac} is not satisfied.
	
	\subsection{Angle Feedback Controllers}
	We define the change in storage power output, $\tilde{P}^{ST} = P^{ST} - \bar{P}^{ST}$, as the difference from the storage power output before a fault occurs. The swing equation of the SMIB system with a local storage device can be rewritten in terms of the angle deviation $\tilde{\delta}$ specifically:
	\begin{equation}\label{eq:storage_smib_deviation_damp}
		M\ddot{\tilde{\delta}} + D\dot{\tilde{\delta}} = \tilde{P}^{ST} + \bar{P}^{E}_{\max}\sin\bar{\delta} - \bar{P}_{\max}^{E}\sin(\tilde{\delta} + \bar{\delta}) ,
	\end{equation}
	with initial angle deviation $\tilde{\delta}_{0} = \delta_{0} - \bar{\delta} \neq 0$. 
	\medskip
	
	In what follows, we present a battery-based control framework considering two different actuator scenarios: one without saturation and the other with a saturation.
	\medskip
	
	{\noindent \bf Battery-based Control without Saturation.} 	Define $x_{3} \in \mathbb{R}$ and $u_{3} \in \mathbb{R}$. Also define $g(u_{3}) = \bar{P}^{E}_{\max}\sin\bar{\delta} - \bar{P}_{\max}^{E}\sin(u_{3} + \bar{\delta})$. The angle feedback controller without actuator saturation is represented by
	\begin{subequations}
		\label{sys:control_without_saturation}
		\begin{align}
			H_{3} : \quad	\dot{x}_{3} &= 
				-\frac{1}{\tau}x_{3} + \frac{K}{\tau}u_{3}, \\
			y_{3} 
			&= x_{3} - Lu_{3}, 
		\end{align}
	\end{subequations}
	where $\tau >0, K >0$, and $L >0.$ Note that this will correspond to a phase-lead compensator.
	\medskip
	
	The following theorem guides us how to select the design choice $(K,L)$. 
	\medskip
	
	\begin{theorem}\label{thm:smib_control_without_saturation}
		Consider the system $H_{1}$ with state-space model~\eqref{sys:smib_reformulate_with_damp} and the system $H_{3}$ with state-space model~\eqref{sys:control_without_saturation}. {\color{red}Then, the equilibrium $(x_{1}^{\ast},x_{3}^{\ast}) = (0,0,0)$ of the feedback  system $(H_{1}, H_{3})$ is globally asymptotically stable, if $K - L < 0$. }
	\end{theorem}
	\medskip
	{\it Proof.} It is straightforward to verify both systems $H_{1}$ and $H_{3}$ are NI with storage functions $V_{1}(x_{1}) = x_{1}^{\top}\begin{bmatrix}
		\frac{M}{2} & 0\\
		0 & 0
	\end{bmatrix} x_{1}$ and $V_{3}(x_{3}) = \frac{1}{2K}x_{3}^{2}$. Then,  we investigate the storage function of the feedback system.  For all $(x_{1}, x_{3}) \in \mathbb{R}^{3} \setminus \{(0,0,0)\}$, we have
	\begin{align*}
		W(x_{1},x_{3}) = &V_{1}(x_{1}) + V_{2}(x_{2}) - \int_{0}^{h_{1}(x_{1})} \big(x_{3} - L\xi\big)d\xi\\
		=&  x_{1}^{\top}\begin{bmatrix}
			\frac{M}{2} & 0\\
			0 & 0
		\end{bmatrix} x_{1} + \frac{x_{3}^{2}}{2K} - x_{3}h_{1}(x_{1}) + \frac{Lh_{1}^{2}(x_{1})}{2}\\
		= & \frac{1}{2} [x_{1}^{\top} \ x_{3} ]Q_{1} [x_{1}^{\top} \ x_{3} ]^{\top},
	\end{align*}
	where $Q_{1} = 
	\begin{bmatrix}
		M &  0 & 0\\
		 0 & L& -1\\
		 0 &  -1 & \frac{1}{K}
	\end{bmatrix} > 0 
	$ when $K - L < 0.$
	We further analyze  the time derivative of the candidate Lyapunov function:
	\begin{align*}
		\dot{W}(x_{1},x_{3}) = [x_{1}^{\top} \ x_{3} ]Q_{2} [x_{1}^{\top} \ x_{3} ]^{\top},
	\end{align*}
	where $Q_{2} = 
	\begin{bmatrix}
		-D &  0 & 0\\
		0 & -\frac{K}{\tau}& \frac{1}{\tau}\\
		0 &  \frac{1}{\tau} & -\frac{1}{\tau K}
	\end{bmatrix} < 0. 
	$ Hence, Theorem~\ref{thm:fb_ni} and Lemma~\ref{lemma:lyapunov_direct_method} are applied to conclude that the equilibrium $(x_{1}^{\ast},x_{3}^{\ast}) = (0,0,0)$ of the  feedback system $(H_{1}, H_{3})$ is locally asymptotically stable.
	Moreover, since the closed-loop linear time-invariant system $(H_{1}, H_{3})$ has only one equilibrium, the equilibrium $(x_{1}^{\ast},x_{3}^{\ast}) = (0,0,0)$ is thus globally  asymptotically stable. The proof is now complete. \hfill$\square$
	\medskip
	
	{\noindent	\textbf{Battery-Based Control with Saturation.}}
	In practical scenarios, the change in storage power output $\tilde{P}^{ST}$ available for sustaining transient stability is constrained by a saturation parameter $b \geq 0$, i.e., $|\tilde{P}^{ST}| < b$. 
	
	Drawing inspiration from Section~\ref{sec:saturated_theory}, the form of the angle feedback controller with  actuator saturation is given by
	\begin{subequations}
		\label{sys:control_with_saturation}
		\begin{align}
			\hat{H}_{3} : \quad \dot{\hat{x}}_{3} &= 
			\begin{cases}
				-\frac{1}{\tau}\hat{x}_{3} + \frac{K}{\tau}\hat{u}_{3}, & |\hat{x}_{3} - L\hat{u}_{3}- g(\hat{u}_{3})| < b,\\
				-\frac{\alpha}{K}\hat{x}_{3}, & |\hat{x}_{3} - L\hat{u}_{3}- g(\hat{u}_{3})| \geq b,
			\end{cases} \\
			\hat{y}_{3} &= g(\hat{u}_{3}) + \text{sat}_{b}(\hat{x}_{3} - L\hat{u}_{3}- g(\hat{u}_{3})),
		\end{align}
	\end{subequations}
	where $\tau >0$, $K >0$, $L >0$ and $\alpha > 0.$
	
	Our following exploration involves two extreme conditions, namely, $b = 0$ and $b = \infty$.
	\begin{itemize}
		\item When $b = 0$: The output of the system $\hat{H}_{3}$ remains unaffected by the state $\hat{x}_{3}$, resulting in $\hat{y}_{3} = g(\hat{u}_{3})$. This scenario corresponds to the absence of a local storage device at the generator bus, without any implemented controller to enhance transient stability. The invariant set is described by the set~\eqref{eq:Region of Attraction}.
		
		\item When $b = \infty$: The controller is not subject to saturation, leading to the system $\hat{H}_{3}$ reducing to the system $H_{3}$. Here, the closed-loop system becomes globally asymptotically stable.
	\end{itemize}
	\medskip
	
	\begin{remark}\label{rmk:battery_based_saturation}
		For $0 < b < \infty$, the feedback system $(H_{1}, \hat{H}_{3})$ is locally stable in the sense of Lyapunov and the invariant set should be larger than the set~\eqref{eq:Region of Attraction}  by continuity. The analytical estimation of the invariant set for the initial state $(\dot{\tilde \delta}_{0}, \tilde{\delta}_{0})$ in the presence of saturation is a complicated task. Hence, we rely on simulations to illustrate the effectiveness of the proposed controller operating with actuator saturation.
	\end{remark}

	\section{Simulations}~\label{sec:simulation}
	Consider an SMIB system with per-unit settings $H = 4$, $D = 0$, $P^{M} = 0.8$, and $P^{E} = 1$. These values are consistent with the numerical example given in \cite{glover2012power}. The nominal system frequency is $50$Hz and the angle at steady state before the occurrence of a fault is $\bar{\delta} = 0.927$ rad. We then consider the occurrence of a fault that cause a new initial angle $\delta(0) = 0.2$ rad.
	\medskip
	
	{\noindent \bf Controller Setting.} The parameters for the angle feedback controller~\eqref{sys:control_with_saturation} are set as $\tau = 0.1, K = 1, L = 1.1,$ and $\alpha = 1.$ We can check that $K - L<0.$ We test three actuator saturation cases $b \in \{0.2, 0.3, \infty\}.$
	\medskip
	
\begin{figure}[tb]
		\centering
		\includegraphics[width=0.8\linewidth]{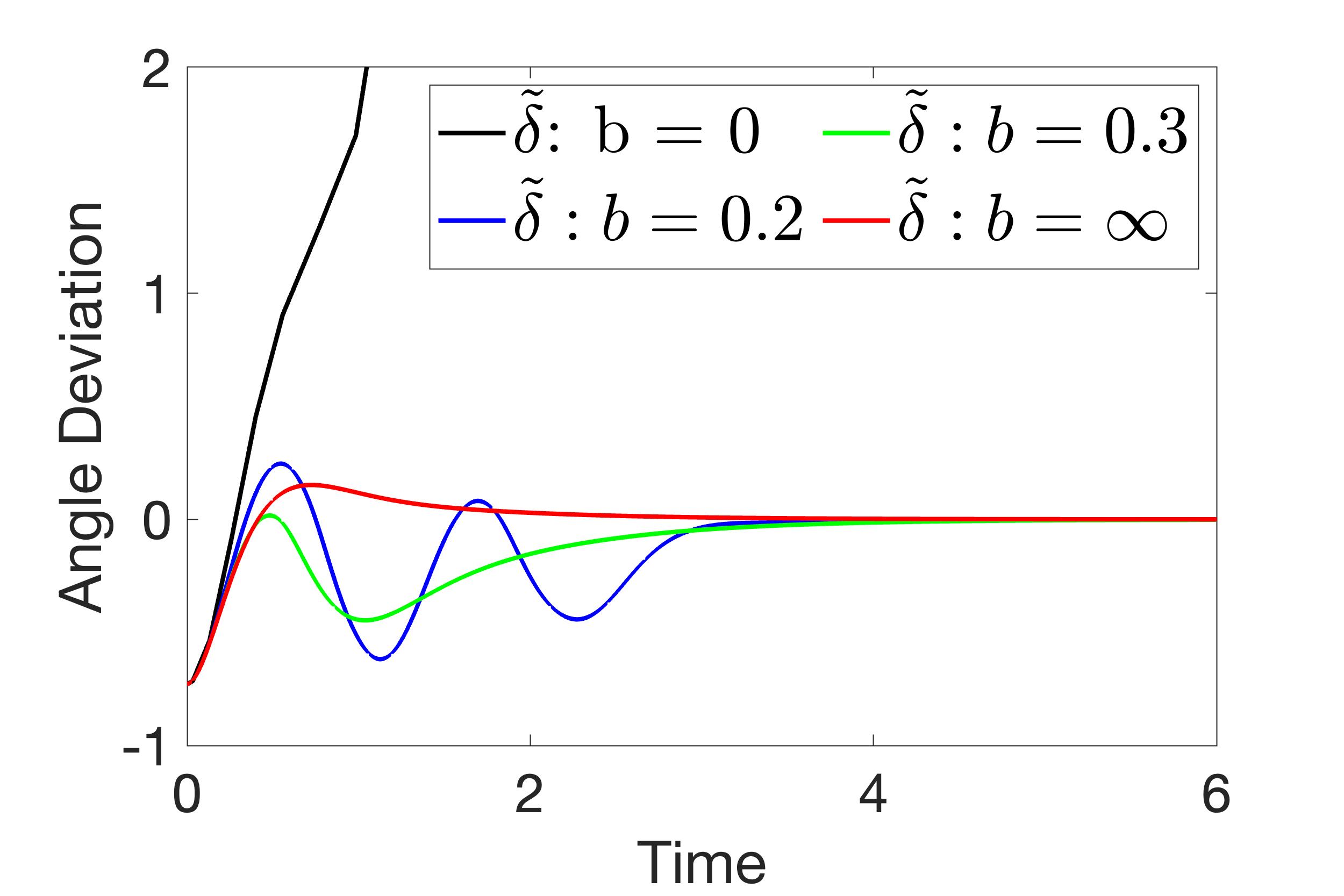}
		\caption{The angle deviation without a controller $b=0$ and with  controllers under three actuator saturation cases $b \in \{0.2, 0.3, \infty\}.$}
		\label{fig:phase_deviation_control}
	\end{figure}

\begin{figure}[tb]
		\centering
		\includegraphics[width=0.8\linewidth]{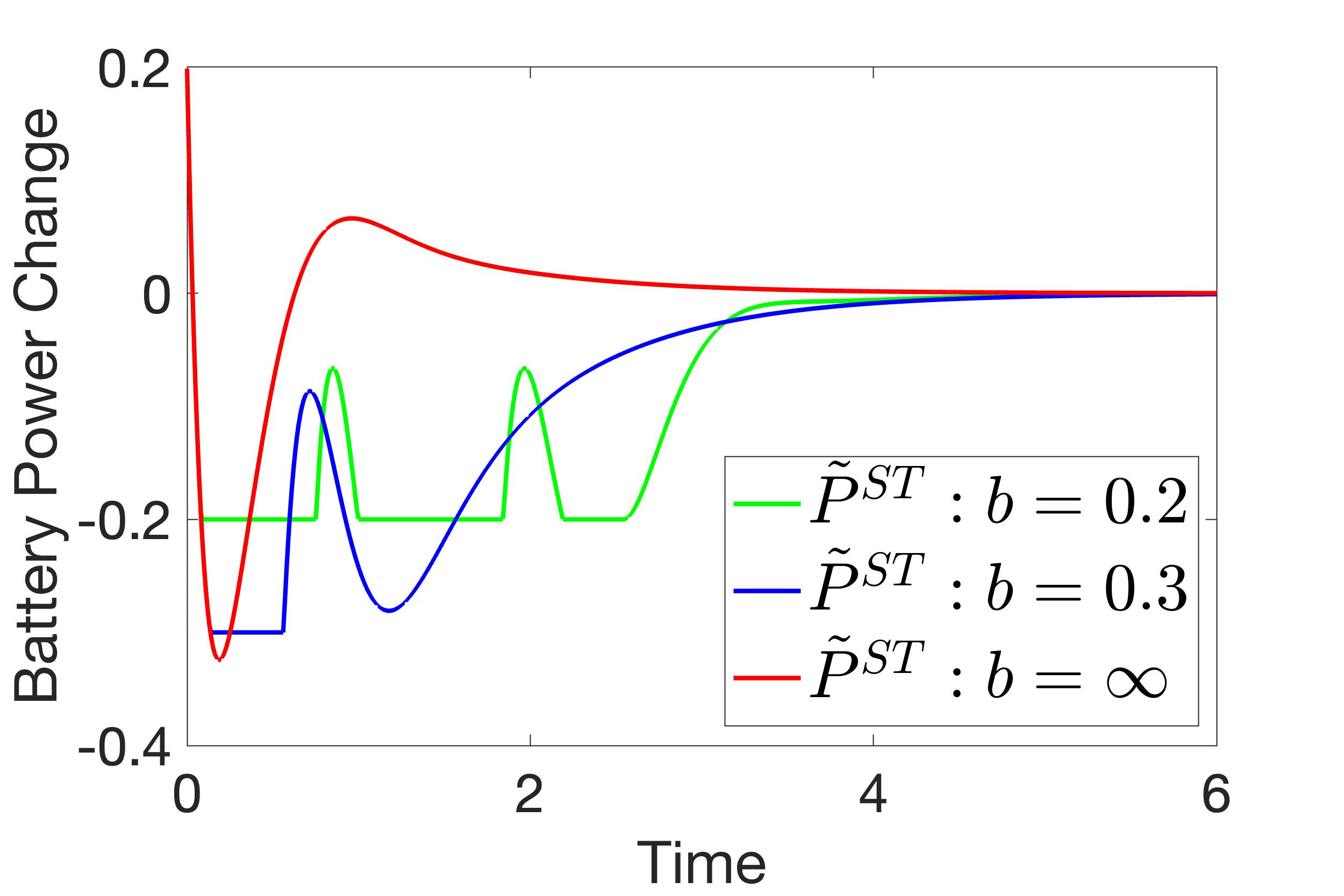}
		\caption{The trajectories of battery power output change under three actuator saturation cases $b \in \{0.2, 0.3, \infty\}.$}
		\label{fig:battery_power}
	\end{figure}

	{\noindent \bf Simulation Results.}  In Fig.~\ref{fig:phase_deviation_control}, we plot the angle deviation for the SMIB system  without a controller $b=0$ and with  controllers under three actuator saturation cases $b \in \{0.2, 0.3, \infty\}$. Since the stability condition of the equal area condition~\eqref{eq:eac} is not satisfied, the SMIB system without a controller loses stability. However, the application of battery-based angle feedback control exhibits its ability in  enhancing transient stability, which validates Theorem~\ref{thm:fb_ni_hat_h3}, Theorem~\ref{thm:smib_control_without_saturation} and Remark~\ref{rmk:battery_based_saturation}. In Fig.~\ref{fig:battery_power}, we plot the trajectories of battery power output change under three actuator saturation cases. Our novel control design is effective in managing the actuator saturation issue, which is consistent with Remark~\ref{rmk:advantage_compensator}.

	\section{Conclusion}\label{sec:conclusion}

	In this paper, we proposed  a nonlinear negative imaginary systems framework for control of electrical power systems. Through the development of a new Lyapunov function fashioned in a manner reminiscent of the Lur\'e-Postnikov form, we established a stability analysis for the feedback interconnection of a nonlinear negative imaginary system and a nonlinear negative imaginary controller. Additionally, we introduced a new class of nonlinear negative imaginary controllers to manage actuator saturation. Our results demonstrated that within this control framework, the controller eventually moved beyond the saturation boundary and the feedback system is locally stable in the sense of Lyapunov. This served as theoretical background supporting the application of battery-based control in electrical power systems. Validation through simulation results for single-machine-infinite-bus power systems further confirmed our results. Investigating large-scale power transmission networks will be a future research direction.

	\bibliographystyle{IEEEtran}
	\bibliography{ref}

\end{document}